\documentclass[twocolumn]{aastex62}

\usepackage{natbib}
\usepackage{amsmath}
\usepackage{textcomp}
\usepackage{graphicx}
\usepackage{ifpdf}

\newcommand{\msun}{$M_{\odot}$}

\newcommand{\midline}{\, | \,}
\newcommand{\hi}{H$\,\textsc{i}$}

\begin{document}

\shorttitle{LMC Ring-like Overdensity}
\title{SMASHing the LMC: Mapping a Ring-like Stellar Overdensity in the LMC Disk}

\correspondingauthor{Yumi Choi}
\email{ymchoi@email.arizona.edu}

\author{Yumi Choi}
\affiliation{Steward Observatory, University of Arizona, 933 North Cherry Avenue, Tucson, AZ 85721, USA}
\affiliation{Department of Physics, Montana State University, P.O. Box 173840, Bozeman, MT 59717-3840, USA}

\author{David L. Nidever}
\affiliation{Department of Physics, Montana State University, P.O. Box 173840, Bozeman, MT 59717-3840, USA}
\affiliation{National Optical Astronomy Observatory, 950 North Cherry Avenue, Tucson, AZ 85719, USA}

\author{Knut Olsen}
\affiliation{National Optical Astronomy Observatory, 950 North Cherry Avenue, Tucson, AZ 85719, USA}

\author{Gurtina Besla}
\affiliation{Steward Observatory, University of Arizona, 933 North Cherry Avenue, Tucson, AZ 85721, USA}

\author{Robert D. Blum}
\affiliation{National Optical Astronomy Observatory, 950 North Cherry Avenue, Tucson, AZ 85719, USA}

\author{Dennis Zaritsky}
\affiliation{Steward Observatory, University of Arizona, 933 North Cherry Avenue, Tucson, AZ 85721, USA}

\author{Maria-Rosa L. Cioni}
\affiliation{Leibniz-Institut f\"{u}r Astrophysics Potsdam (AIP), An der Sternwarte 16, D-14482 Potsdam, Germany}

\author{Roeland P. van der Marel}
\affiliation{Space Telescope Science Institute, 3700 San Martin Drive, Baltimore, MD 21218, USA}
\affiliation{Center for Astrophysical Sciences, Department of Physics \& Astronomy, Johns Hopkins University, Baltimore, MD 21218, USA}

\author{Eric F. Bell}
\affiliation{Department of Astronomy, University of Michigan, 1085 S. University Avenue, Ann Arbor, MI 48109-1107, USA}

\author{L. Clifton Johnson}
\affiliation{Department of Physics and Astronomy, Northwestern University, 2145 Sheridan Road, Evanston, IL 60208, USA}

\author{A. Katherina Vivas}
\affiliation{Cerro Tololo Inter-American Observatory, National Optical Astronomy Observatory, Casilla 603, La Serena, Chile}

\author{Alistair R. Walker}
\affiliation{Cerro Tololo Inter-American Observatory, National Optical Astronomy Observatory, Casilla 603, La Serena, Chile}

\author{Thomas J. L. de Boer}
\affiliation{Department of Physics, University of Surrey, Guildford, GU2 7XH, UK}

\author{Noelia E. D. No\"el}
\affiliation{Department of Physics, University of Surrey, Guildford, GU2 7XH, UK}

\author{Antonela Monachesi}
\affiliation{Instituto de Investigaci\'on Multidisciplinario en Ciencias y Tecnolog\'ia, Universidad de La Serena, Ra\'ul Bitr\'an 1305, La Serena, Chile}
\affiliation{Departamento de F\'isica y Astronom\'ia, Universidad de La Serena, Av. Juan Cisternas 1200 Norte, La Serena, Chile}

\author{Carme Gallart}
\affiliation{Instituto de Astrof\'{i}sica de Canarias, La Laguna, Tenerife, Spain}

\author{Matteo Monelli}
\affiliation{Instituto de Astrof\'{i}sica de Canarias, La Laguna, Tenerife, Spain}
\affiliation{Departamento de Astrof\'{i}sica, Universidad de La Laguna, Tenerife, Spain}

\author{Guy S. Stringfellow}
\affiliation{Center for Astrophysics and Space Astronomy, University of Colorado, 389 UCB, Boulder, CO 80309-0389, USA}

\author{Pol Massana}
\affiliation{Department of Physics, University of Surrey, Guildford, GU2 7XH, UK}

\author{David Martinez-Delgado}
\affiliation{Astronomisches Rechen-Institut, Zentrum f\"ur Astronomie der Universit\"at Heidelberg,  M{\"o}nchhofstr. 12-14, D-69120 Heidelberg, Germany}

\author{Ricardo R. Mu\~noz}
\affiliation{Departamento de Astronom\'ia, Universidad de Chile, Camino del Observatorio 1515, Las Condes, Santiago, Chile}

\shortauthors{Choi et al.}

\begin{abstract}
We explore the stellar structure of the Large Magellanic Cloud (LMC) disk using data from the Survey of the MAgellanic Stellar History (SMASH) and the Dark Energy Survey. We detect a ring-like stellar overdensity in the red clump star count map at a radius of $\sim$6\degr~($\sim$5.2~kpc at the LMC distance) that is continuous over $\sim$270\degr~in position angle and is only limited by the current data coverage. The overdensity shows an amplitude up to 2.5 times higher than that of the underlying smooth disk. This structure might be related to the multiple arms found by de Vaucouleurs. We find that the overdensity shows spatial correlation with intermediate-age star clusters, but not with young ($<$ 1~Gyr) main-sequence stars, indicating the stellar populations associated with the overdensity are intermediate in age or older. Our findings on the LMC overdensity can be explained by either of two distinct formation mechanisms of a ring-like overdensity: (1) the overdensity formed out of an asymmetric one-armed spiral wrapping around the LMC main body, which is induced by repeated encounters with the Small Magellanic Cloud (SMC) over the last Gyr, or (2) the overdensity formed very recently as a tidal response to a direct collision with the SMC.  Although the measured properties of the overdensity alone cannot distinguish between the two candidate scenarios, the consistency with both scenarios suggests that the ring-like overdensity is likely a product of tidal interaction with the SMC, but not with the Milky Way halo. 
\end{abstract}

\keywords{galaxies: dwarf --- galaxies: interactions --- galaxies: Magellanic Clouds --- galaxies: structure}

\section{Introduction} \label{sec:intro}
Contrary to a long-held view that the Milky Way (MW) is the main perturber of the Large Magellanic Cloud \citep[LMC; e.g.,][]{fujimoto77,knunkel79,murai80,lin82,gardiner94,bekki05,diaz12}, over the last decade increasing evidence has suggested that the Small Magellanic Cloud (SMC) is instead responsible for most of the perturbations \citep[e.g.,][]{besla12,besla16}. According to high-precision proper motions from the Hubble Space Telescope \citep{kallivayalil06b,kallivayalil06a,piatek08,kallivayalil13} and confirmed by \citet{vandermarel16} using Gaia DR1 \citep{gaia16}, the LMC--SMC pair (MCs) is likely on its first infall into the MW, whereas the Clouds have been gravitationally bound to each other at least for several Gyr \citep{besla12}. The measurements of the MCs' proper motions using Gaia DR2 are also consistent with these previous results \citep{helmi18}.

The LMC disk as a whole is generally well described as a planar disk, and it exhibits many interesting asymmetric features such as one prominent spiral arm, spur-like structures, an off-centered bar \citep{devaucouleurs55,devaucouleurs72}, an inner warp \citep{olsen02}, an outer warp \citep{choi18}, a shell-like structure \citep{devaucouleurs55,irwin91}, and stellar substructures in the periphery of the disk \citep{mackey16,mackey18}. Many of these structures are common in Magellanic Irregulars \citep{devaucouleurs72}, and can be explained naturally by repeated interactions with their companions \citep[e.g.,][]{athanassoula96,berentzen03,bekki09,besla16}. In particular, the formation of off-centered stellar bars seems to require tidal interactions with smaller halos \citep[e.g.,][]{bekki09,besla12}. 

\citet{besla12} proposed the possibility of a recent direct collision following a number of flyby encounters between the MCs. A number of observational lines of evidence support a direct collision between the MCs a few hundred Myr ago \citep[e.g.,][]{olsen11,noel15,besla16,carrera17,choi18,niederhofer18,zivick18}. For example, \citet{choi18} detected a new warp in the outer southwestern LMC disk as well as a tilted off-centered bar and the inner warp, which was previously found by \citet{olsen02}. \citet{zivick18} reported updated $HST$ proper motions of the SMC and used orbit calculations to find that the mean impact parameter of the MCs collision is 7.5$\pm$2.5~kpc. 

A direct collision by a satellite galaxy can produce significant ripples in the host galaxy \citep[e.g.,][]{gomez13,gomez17}. Coherent ripples in the disk can then appear as a ring-like stellar overdensities \citep{purcell11}. If this has occurred in the LMC as well, then overdensitiies and underdensities across the LMC disk can be expected, either resulting from a direct collision or repeated perturbation by the SMC. 

A shell-like (we call it ``ring-like'') structure in the LMC disk has been mentioned in the literature, but only very briefly. \citet{devaucouleurs55} made the first detection of this structure as a faint outer loop based on his long exposure photographs, and interpreted this structure as a one-arm spiral emanating from the western end of the bar and wrapping around the LMC main body. A hint of this structure was also found in the distribution of star clusters \citep[e.g.,][]{westerlund64,irwin91} and in a ratio map of carbon-rich to oxygen-rich asymptotic giant branch (AGB) stars \citep{cioni03}. According to the more recent star cluster catalog by \citet{bica08}, this structure is outlined by the spatial distribution of intermediate-age star clusters, older than stellar associations ($>$ a few Myr) and younger than old ($<$ 4~Gyr) star clusters (see their Figure 2). 

Evidence of this ring structure can be also found in more recent stellar number density studies. For example, \citet{vandermarel01b} found an excess of stars in their 1D radial stellar number density profile for red giant branch (RGB) and AGB stars from the Two Micron All Sky Survey \citep{skrutskie06} and the Deep Near-Infrared Southern Sky Survey \citep{epchtein97} data. They suggested a tidal origin for this excess. An impression of the ring-like stellar structure is also present in the Gaia DR1 and DR2 star count maps in \citet{belokurov17} and \citet{helmi18}, respectively. Because the emphases of their work were on the stellar tidal tails at larger radii around the MCs using variable stars \citep{belokurov17} and on the analysis of the stellar proper motions \citep{helmi18}, nothing was mentioned about the ring-like structure that is the focus of our study.
  
In short, while this stellar overdensity has been known to exist for over 60 years, no study has conducted a detailed investigation to characterize the structure quantitatively and understand its origin. In this paper, we firmly establish the existence of the ring-like overdensity in the LMC, map the structure via detailed disk modeling, characterize its properties quantitatively for the first time, and discuss its possible origin. Understanding the origin of this structure is crucial to our understanding of the interaction history of the LMC--SMC system, and will provide insights into the evolution of interacting Magellanic-type galaxies in general.

\section{Data} \label{sec:data}
\subsection{Survey of the MAgellanic Stellar History (SMASH) Data}
The SMASH is a Dark Energy Camera \citep[DECam;][]{flaugher15} survey that has completely mapped the main bodies of the Clouds, and probed the Magellanic Periphery and the Leading Arm region by sampling ``island" fields over a $\sim$2400~deg$^2$ area of the sky, at an $\sim$20\% filling factor. The goal of the deep photometric ($ugriz$$\sim$24th mag) survey is to study the stellar structure and star formation history of the Clouds, with particular focus on low density stellar populations associated with the stellar halos and tidal debris of the Magellanic Clouds.  Using old, main-sequence (MS) stars as a tracer, a technique previously adopted by the Outer Limits Survey \citep{Saha2010}, SMASH data should be able to reveal the relics of the formation and past interactions of the MCs down to surface brightnesses equivalent to $\Sigma_{g}=$ 35~mag~arcsec$^{-2}$ over a vast area.

\citet{nidever17} described the reduction and calibration of the data as well as the first public data release, which contains $\sim$700 million measurements of $\sim$100 million objects in 61 deep and fully calibrated fields from the SMASH survey\footnote{http://datalab.noao.edu/smash/smash.php}. The photometric precisions of the final SMASH catalogs are roughly 1.0\% ($u$), 0.7\% ($g$), 0.5\% ($r$), 0.8\% ($i$), and 0.5\% ($z$). The obtained calibration accuracies are 1.3\% ($u$), 1.3\% ($g$), 1.0\% ($r$), 1.2\% ($i$), and 1.3\% ($z$).  The median 5$\sigma$ point-source depths in the $ugriz$ bands are (23.9, 24.8, 24.5, 24.2, 23.5) mag. The astrometric precision is $\sim$15~mas and accuracy is $\sim$2~mas with respect to the Gaia DR1 astrometric reference frame \citep{gaia16}. We refer the readers to \citet{nidever17} for further details on the survey, data reduction, and photometry. 

In this study, we use 62 fields\footnote{We note that only two (Field 44 and Field 55) of these 62 fields were included in the first data release. SMASH DR2 will be available in early 2019.} that cover the LMC main body, and use the red clump (RC) and bright MS stars to study the LMC's structure. In each field catalog, we first select point-like sources using the DAOPHOT morphology parameters ``chi'' and ``sharp'': chi$<$3 ($<$5 for Fields 36 and 41 in the center) and $\midline$sharp$\midline$$<$1 ($<$2 for Fields 36 and 41). Contamination by unresolved background galaxies and the foreground MW stars inside the LMC main body is negligible, especially within the magnitude range we are interested in.

\begin{figure*}
\centering
\hspace{-0.5cm}
\includegraphics[trim=0cm 0cm 1cm 0cm, clip=True, width=9cm]{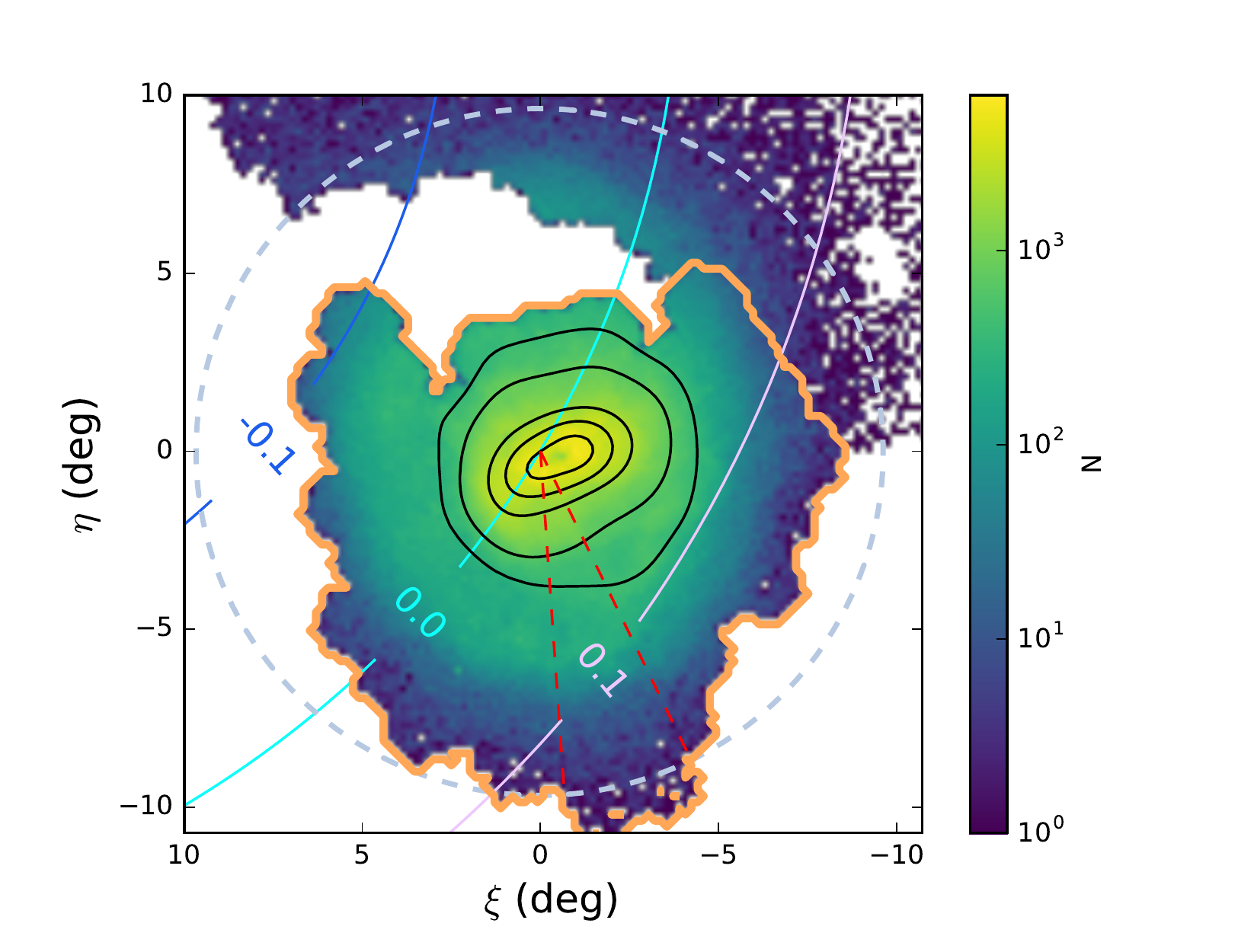}
\includegraphics[trim=0cm 0cm 1cm 1cm, clip=True, width=9cm]{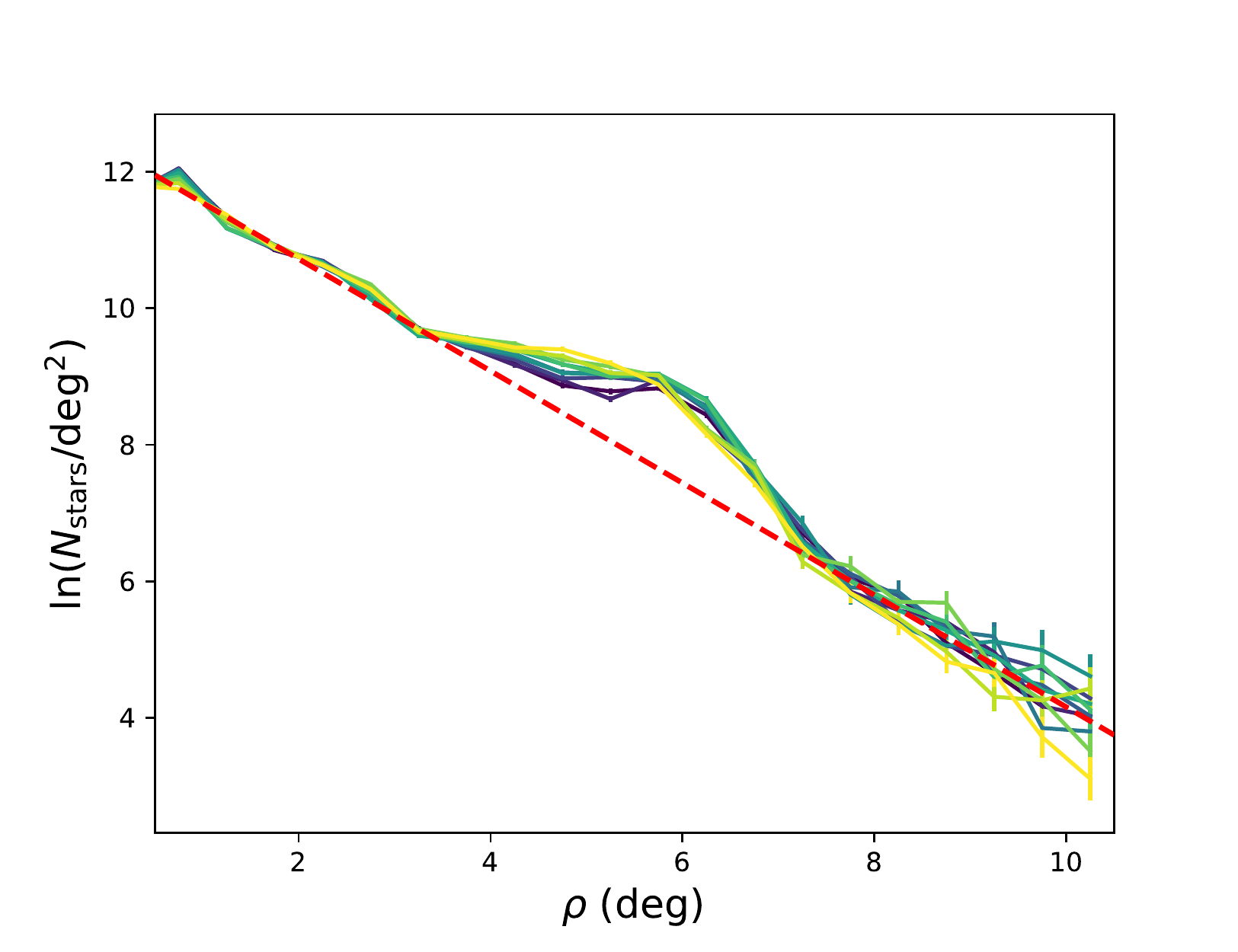}
\caption{Left: Density map of SMASH and DES selected RC stars. North is up and east is left. The region contiguously covered by SMASH data in the LMC disk is outlined in orange.  The three curved lines indicate magnitude offsets of $-$0.1 mag, 0.0 mag, and $+$0.1 mag in distance modulus due to the inclination of the LMC disk. The dashed circle is at a radius of 9.5\degr. We also overplot the density contours in the inner disk to better show the shape of the bar. The transition of the contour shapes from bar-dominated to disk-dominated is clearly seen at 2\degr\,$< \rho <$ 4\degr. Right: 11 RC number density radial profiles in the sector denoted as red dashed lines in the left panel. This sector is chosen due to the most complete radial coverage in our RC map and the minimum contribution of the bar to the radial profile. We divide the sector into 11 slices in azimuthal angle of 2\degr\,and compute stellar density radial profiles in each slice. All profiles are well described as a single exponential disk with a prominent excess peaked around 6\degr\,from the LMC center.
\label{rcmap}}
\end{figure*}

\subsection{DES Data}
The Dark Energy Survey Data Release 1 (DES DR1) was made earlier this year \citep{desdr1}, and it is easily accessible through the NOAO Data Lab\footnote{http://datalab.noao.edu/index.php}. In addition to the SMASH data, we use DES data near the LMC main body to better constrain the exponential disk by adding coverage of the northern disk up to $\rho$ = 9.5\degr\,from the center (see Figure~\ref{rcmap}). Beyond $\sim$9.5\degr\,to the north, the number density of LMC RC stars decreases rapidly and the level of contamination from MW foreground becomes severe. Thus, we only make use of the DES data inside $\rho$ = 9.5\degr\,for modeling the LMC disk.

\section{Spatial Distribution of Various Stellar Populations}
The geometric distributions of young and old populations are significantly different from each other in the LMC disk \citep[e.g.,][]{zaritsky04a,balbinot15}. In this section, we describe our selection of the RC stars, which are a representation of an intermediate-age population, and bright MS stars, which are a representation of an young population, and then briefly discuss their spatial distribution. The RC is the stellar population used in \citet{choi18} to measure the LMC disk's inclination and the line-of-nodes position angle that we adopt in this study when modeling the LMC RC disk. We use the bright MS population to get a rough constraint on the ages of stars composing the ring-like overdensity that we investigate. 

\subsection{Selecting Red Clump Stars}
The RC is one of the most prominent features in the LMC color-magnitude diagram (CMD). RC stars are in the core He-burning stage, and have relatively low masses with intermediate ages on average \citep[$\lesssim$2~Gyr;][]{castellani00} and moderately high metallicity. In the CMD, RC stars occupy a well-defined, narrow region due to their uniform core mass, regardless of their initial mass. This fundamental property results in very limited effective temperatures and luminosities, allowing RC stars to be used to accurately measure distances and extinctions \citep[and references therein]{girardi16}.  

We follow the procedure used in \citet{choi18} to select LMC RC stars. Briefly, we carefully define the RC selection box rather than using a simple rectangular region around the RC \citep[see Figure~3 in][]{choi18} to minimize the contaminants in the RC selection. To do that, we define a large initial box around the RC with a $g$-band magnitude range between 18.5 and 21. The color range varies for each field according to its dust extinction, while the slope is fixed along the extinction vector. The maximum color is set by the contamination by MW foreground stars. The number of unique SMASH LMC RC stars is $\sim$2.2~million.  

The RC selection from the DES data is nearly identical to that used for the SMASH data. One difference is that we select the RC field-by-field from the SMASH data where dust and variations in stellar population govern the RC morphology, while we select the RC from the DES data in four azimuthal slices to have enough RC stars to be clumped in a CMD and to minimize possible contamination at the same time. The RC morphology does not change very much in the outer LMC disk due to rather homogeneous populations and lack of internal dust, but the shift in the magnitude with position angle due to the disk inclination is not negligible. Thus, we slice the DES data azimuthally to limit the shift in the RC magnitude by 0.1~mag based on the predicted magnitude variation map generated with the inclination and line-of-nodes measurements from Choi et al. (2018). This allows us to reduce the contamination from non-RC stars by keeping the RC selection box fairly narrow. The solid arcs in Figure~\ref{rcmap} show the predicted magnitude variation with respect to the fiducial magnitude. The LMC line of nodes is the tangent line to the $\Delta\,m$ = 0~mag arc at the LMC center.

The left panel of Figure~\ref{rcmap} shows the star count map of the selected RC stars from both the SMASH and DES data. There is a smooth continuation between the two data sets. The orange line outlines the region contiguously covered by the SMASH data. The combined SMASH and DES area enables us to encompass a sufficient portion of the LMC disk to cover an additional $\sim$4\degr\,beyond the ring-like structure detected in the literature, particularly the northern and southern parts. We refer the readers to \citet{choi18} for a more detailed discussion on the intrinsic RC properties, our RC selection in CMDs, and the resulting RC star count map. In the right panel, we present the 1D RC stellar density radial profiles for 11 azimuthal bins with a 2\degr\,width within the sector denoted in the left panel. This sector covers the largest radial extent (up to 10.5\degr) in our map. Because the direction of the sector is roughly parallel to the bar minor axis, the contribution of the bar to the radial profile is limited to a small radius ($<$1.5\degr) in this sector. All 11 radial profiles are well described as a single exponential disk (red dashed-line), showing an excess of RC stars at around 6\degr\,from the LMC center. \citet{majewski09} and \citet{nidever18} also showed that the LMC density radial profile is best matched by the exponential disk profile out to $\sim$9\degr\,and $\sim$13\degr, respectively. This, combined with the sufficient coverage of the LMC disk, gives us confidence that our data coverage will allow reliable characterization of the overdensity.

\subsection{Selecting Young Stellar Populations}
\begin{figure*}
\centering
\includegraphics[trim=1cm 10cm 0cm 2.5cm,clip=True,width=20cm]{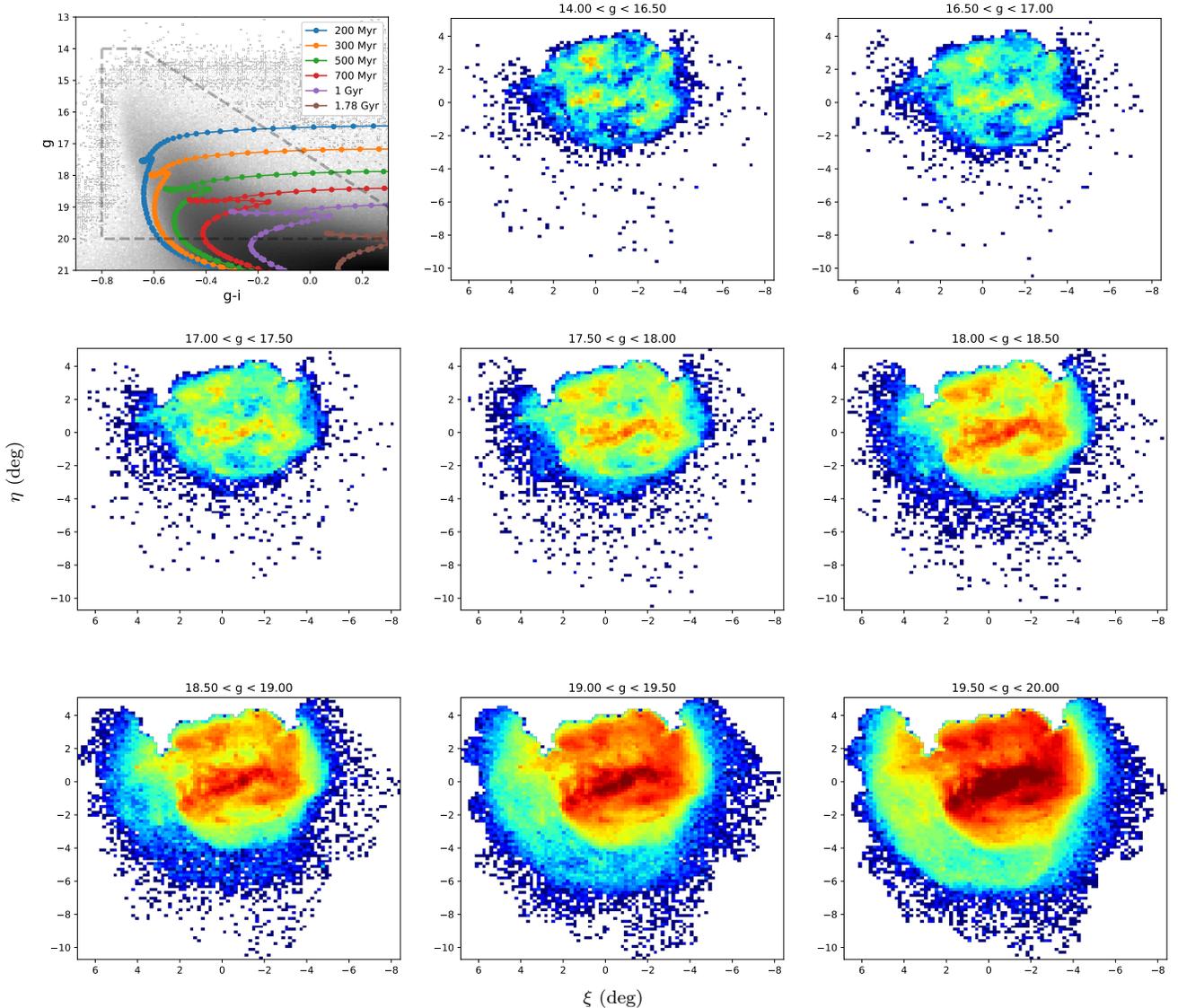}
\caption{Top Left: Combined Hess diagram of SMASH LMC bright MS stars. The dashed polygon indicates the region used to select young main-sequence stars. Five isochrones of different ages with $Z$ = 0.012 \citep[PARSEC;][]{bressan12,marigo17} are also shown. Rest: Spatial distribution of bright MS stars in magnitude bins. The color-map scale remains the same across all panels. Each panel contains stellar populations with a maximum age that increases with increasing magnitude. The youngest stars are mostly found in the central region, while older stars tend to trace the extended disk. The faintest two bins corresponding to maximum ages of $\sim$1.5--1.8 Gyr start showing a feature consistent with our ring-like overdensity seen in the RC map.
\label{yMSmap}}
\end{figure*}

The top left panel of Figure \ref{yMSmap} shows the combined Hess diagram of SMASH LMC bright MS stars. For bright MS, we select stars with $g$-band magnitudes brighter than 20~mag and -0.8 $< g-i <$ 0.3. The gray dashed polygon denotes our selection box. The number of selected unique bright MS stars is $\sim$3~million. We do not include the DES data to construct a list of bright MS stars because the SMASH data alone provide sufficient information on the spatial distribution of bright MS stars and there is no significant number of bright MS stars in the DES region. 

Figure \ref{yMSmap} shows the spatial distribution of selected bright MS stars in a range of magnitude bins. The maximum age in each bin increases from $\sim$100~Myr to $\sim$1.8~Gyr. While the RC stars show a relatively smooth and extended spatial distribution, the young MS stars are highly structured and mostly confined to the inner disk. The MS stars in brighter magnitude bins ($g <$ 17.5~mag) are distributed in clumpy structures, while fainter ones (17.5 $< g <$ 19~mag) form coherent structures such as the central bar and one prominent spiral arm, indicating hierarchical star formation \citep[e.g.,][]{harris09,sun17}. However, the latter structures are only evident when looking at MS stars fainter than $\sim$18~mag. The lack of the bar structure in the youngest bins agrees with the detailed star formation histories of the LMC bar \citep{monteagudo18}. The MS stars fainter than $\sim$19~mag extend to even larger radii and start resembling the RC spatial distribution. The maps of bright MS stars as a function of their apparent magnitude suggest that the star forming disk has dramatically shrunk over the last $\sim$1--2~Gyr from a radius of $\sim$7\degr\,to $\sim$4\degr. This is consistent with the outside-in LMC evolution scenario \citep{Meschin2014}.

\subsection{Coordinate Definition and Transformation}
In this section, we briefly introduce each of the three coordinate systems used in this work following \citet{vandermarel01a} and \citet{mackey16}: angular coordinates ($\rho$,$\phi$), tangent plane coordinates ($\xi$,$\eta$), and the galaxy plane coordinates ($x$,$y$,$z$). For coordinate transformation from the celestial coordinates (right ascension $\alpha$, declination $\delta$) and distance $D$, we adopt the LMC center as the origin with ($\alpha_0$,$\delta_0$) = (82.25\degr,$-$69.50\degr) and $D_0$ = 49.9~kpc \citep{vandermarel01b,degrijs14}. We adopt the center of the bar as ($\alpha$, $\delta$) = (79.91\degr, $-$69.45\degr) \citep{devaucouleurs72}. In addition, we use inclination and line-of-nodes position angle ($i$,$\theta$) = (25.86\degr,149.23\degr) as determined by \citet{choi18} using the 3D spatial distribution of the RC. 

Angular coordinates ($\rho$,$\phi$) are defined on the celestial sphere around the origin point ($\alpha_0$,$\delta_0$) as follows:
\begin{equation}
\begin{aligned}
  \cos(\rho) &= \sin(\delta)\sin(\delta_0) + \cos(\delta)\cos(\delta_0)\cos(\alpha - \alpha_0) \\
  \tan(\phi) &= \frac{\cos(\delta)\sin(\delta_0)\cos(\alpha - \alpha_0) - \sin(\delta)\cos(\delta_0)}{\cos(\delta)\sin(\alpha - \alpha_0)}. 
\end{aligned}
\end{equation}
where $\rho$ is the angular distance of a point ($\alpha$,$\delta$) from the origin and $\phi$ is the position angle measured counterclockwise from the west at constant decl. $\delta_0$. 

The ($\xi$,$\eta$) tangent plane coordinates are obtained through the gnomonic projection of the sphere onto a tangent plane:  
\begin{equation}
\begin{aligned}
  \xi &= \frac{\cos(\delta)\sin(\alpha-\alpha_{0})}
             {\sin(\delta_{0})\sin(\delta) + \cos(\delta_{0})\cos(\delta)\cos(\alpha-\alpha_{0})} \\
  \eta &= \frac{\cos(\delta_{0})\sin(\delta) - \sin(\delta_{0})\cos(\delta)\cos(\alpha-\alpha_{0})}
             {\sin(\delta_{0})\sin(\delta) + \cos(\delta_{0})\cos(\delta)\cos(\alpha-\alpha_{0})}. 
\end{aligned}
\end{equation}
The observed and modeled stellar densities are compared in 10\arcmin$\times$10\arcmin\,cells in this tangent plane (see Section~\ref{modeldisk}).

A Cartesian coordinate system ($x$,$y$,$z$) is used to represent the LMC disk plane. The ($x$,$y$)-galaxy plane is defined as an infinitely thin plane that is inclined with respect to the sky plane by an angle $i$ about the line of nodes with a position angle $\theta$. We compute the expected stellar number density at positions in the disk plane. The coordinates are  
\begin{equation}
\begin{aligned}
  x &= D\sin(\rho)\cos(\phi - \theta) \\
  y &= D[\sin(\rho)\sin(\phi - \theta)\cos(i) + \cos(\rho)\sin(i)] - D_0\sin(i) \\
  z &= D[\sin(\rho)\sin(\phi - \theta)\sin(i) - \cos(\rho)\cos(i)] + D_0\cos(i). 
\end{aligned}
\end{equation}
In the LMC galaxy plane (i.e., $z$ = 0), the line-of-sight distance ($D$) is  
\begin{equation}
D = \frac{D_0\cos(i)}{[\cos(i)\cos(\rho) - \sin(i)\sin(\rho)\sin(\phi - \theta)]}.
\end{equation}

\section{Decomposing the LMC Disk}
We model the observed RC star count map as a two-component galaxy with an exponential disk and a boxy bar to constrain the underlying smooth disk structure and better characterize the overdense substructure. 

\subsection{The Exponential Disk}
A single, symmetric exponential disk requires four fitting parameters: central stellar number density ($\mu_{0,d}$), disk scale length ($r_d$), axis ratio ($b/a$), and 
semi-major axis position angle ($\psi$):   
\begin{equation}
\label{eq:expdisk} 
\mu_{d}(r)=\mu_{0,d}\,\exp\left(-\frac{r}{r_d}\right), 
\end{equation}
 where $r$ is 
 \begin{equation}
\label{eq:genEll}
r(x,y)^{2} = \left(x\cos(\psi)-y\sin(\psi)\right)^{2}+\left(\frac{x\sin(\psi)+y\cos(\psi)}{b/a}\right)^{2}.
\end{equation}

\subsection{The Off-centered Bar}
In contrast to an exponential disk, it is non-physical to model a bar as an inclined 2D structure because of its significant $z$-direction thickness compared to its $x,y$ dimension. Thus, we model the 2D bar as it appears in the tangent plane. 

For the 2D elliptical bar, there are seven fitting parameters, bringing the total number of parameters included in the two-component modeling to 11. We adopt the modified Ferrer profile for a bar \citep[e.g.,][]{laurikainen07,peng10}, 
\begin{equation}
\label{eq:ferrers} 
\mu_{b}(r)=\mu_{0,b}\,\left[1 - \left(\frac{r}{r_{out}}\right)^{2-\beta}\right]^{\alpha}, 
\end{equation}
where $\mu_{0,b}$ is the bar central number density, $r_{out}$ is the end of the bar profile, $\alpha$ determines the sharpness of the truncation at $r_{out}$, and $\beta$ determines the inner slope of the profile.

To account for the boxiness/diskiness, we apply a generalized ellipse \citep{athanassoula90} as follows: 
\begin{align}
\label{eq:genBar}
r(x,y) =& \Bigg(\left|x\cos(\psi)-y\sin(\psi)\right|^{C_0} \Bigg.\nonumber\\
&~~\Bigg.+\left|\frac{x\sin(\psi)+y\cos(\psi)}{b/a}\right|^{C_0}\Bigg)^{\frac{1}{C_0}}.
\end{align}
The $C_{0}$ parameter determines the shape of the bar: $C_{0} =$ 2 -- purely elliptical; $C_{0} >$ 2 -- boxy;  $C_{0} <$ 2 -- disky. The bar ellipticity can be determined from the $b/a$ ratio, $\epsilon = 1 - (b/a)$. 

The final model prediction in each cell ($ij$) is the sum of the projected disk component in the tangent plane and the bar component: 
\begin{equation}
\begin{aligned}
\label{eq:genSum}
\mu_{m,ij} &= \mu_{d,ij} + \mu_{b,ij} \\
           &= \frac{\cos(\rho)^3}{\cos(\rho\sin(\phi-\theta)+i)}\left(\frac{D}{D_0}\right)^2\mu_{d,xy} + \mu_{b,ij}.
\end{aligned}
\end{equation}
An extra term in front of $\mu_{d,xy}$, which is the exponential disk central number density in the galaxy plane, takes care of the factors from the gnomic projection of the predicted number of stars in a given area on the galaxy plane at a distance $D$ as well as a distance gradient across the given cell in the tangent plane.

\begin{deluxetable*}{lccc}
\tablecaption{LMC Disk Parameters
\label{param_2comp}}
\tablewidth{0pt}
\tablehead{
\colhead{Parameter} & \colhead{Description} &  \colhead{Range for Flat Prior} &\colhead{Results}}
\startdata
$\mu_{0,d}$ [$N_{\rm stars}$/kpc$^{2}$]  & disk central number density & 0--$\infty$ & 2800.89$^{+7.35}_{-7.38}$  \\
$r_{d}$ [kpc]  & disk scale length  & 0--$\infty$ & 1.667$^{+0.002}_{-0.002}$\\ 
$(b/a)_{\rm disk}$   & disk axis ratio  & 0--1  & 0.836$^{+0.001}_{-0.001}$ \\
PA$_{\rm disk}$ [degree]  & disk major axis position angle  & 0--$2\pi$   &  227.24$^{+0.178}_{-0.188}$\\ 
$\mu_{0,b}$ [$N_{\rm stars}$/$''^{2}$]  & bar central number density                          & 0--$\infty$ & 2474.26$^{+14.21}_{-14.73}$ \\
$r_{\rm out}$ [degree]  & bar length   & 0--3.5 & 3.499$^{+0.000}_{-0.000}$ \\
$\alpha$   & sharpness of the truncation of bar profile & 0--10.5     & 1.371$^{+0.012}_{-0.012}$ \\
$\beta$   & inner slope of bar profile    & 0--2 & 0.427$^{+0.021}_{-0.020}$ \\
$C_{0}$  & bar shape parameter  & 1.5--3.0         & 2.999$^{+0.000}_{-0.000}$ \\
$(b/a)_{\rm bar}$  & bar axis ratio  & 0--1          & 0.446$^{+0.001}_{-0.001}$ \\
PA$_{\rm bar}$ [degree]  & bar position angle  & 0--$2\pi$  & 154.18$^{+0.11}_{-0.11}$ \\
\enddata
\end{deluxetable*}

\begin{figure*}
\centering
\includegraphics[width=15cm]{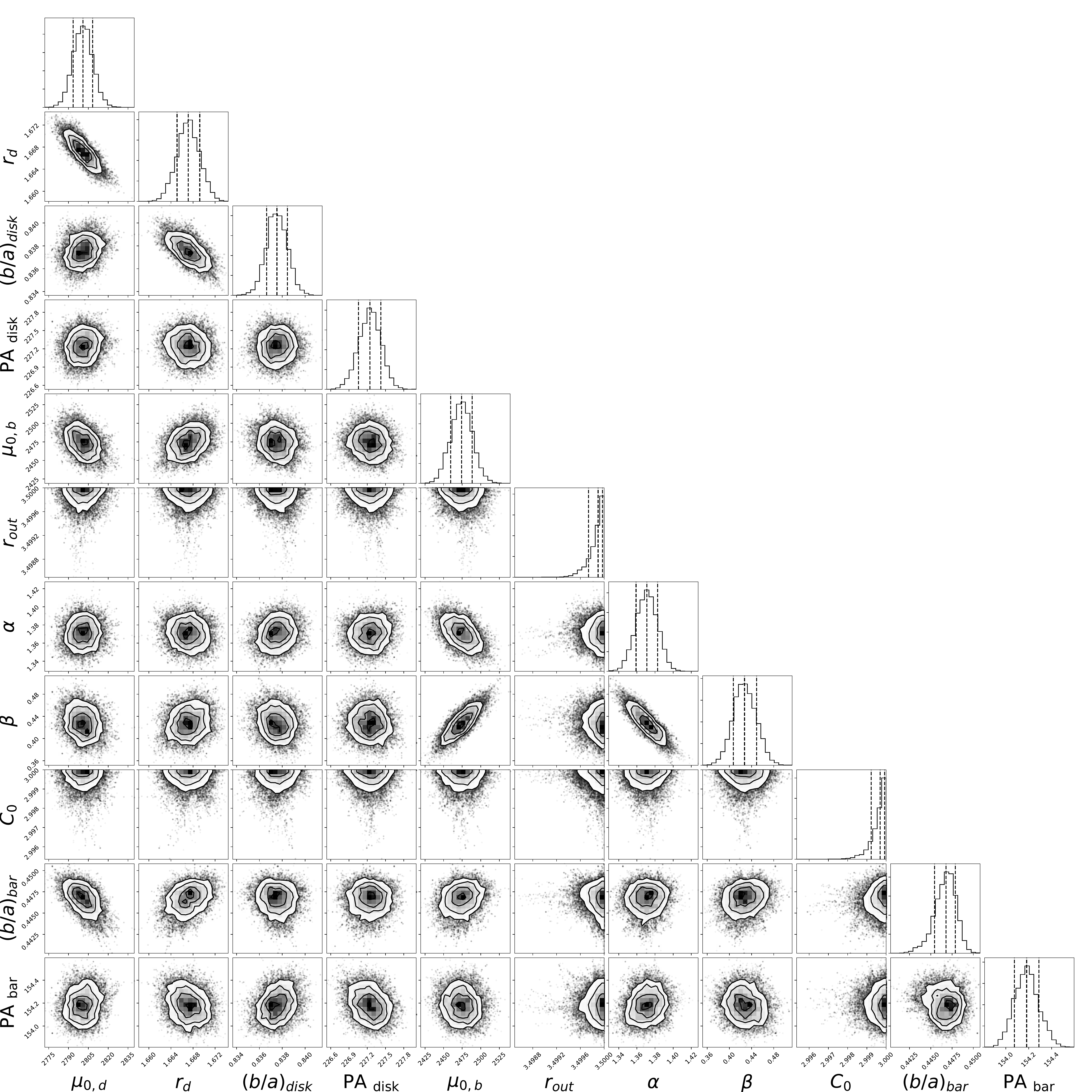}
\llap{\raisebox{9cm}{\includegraphics[width=8cm]{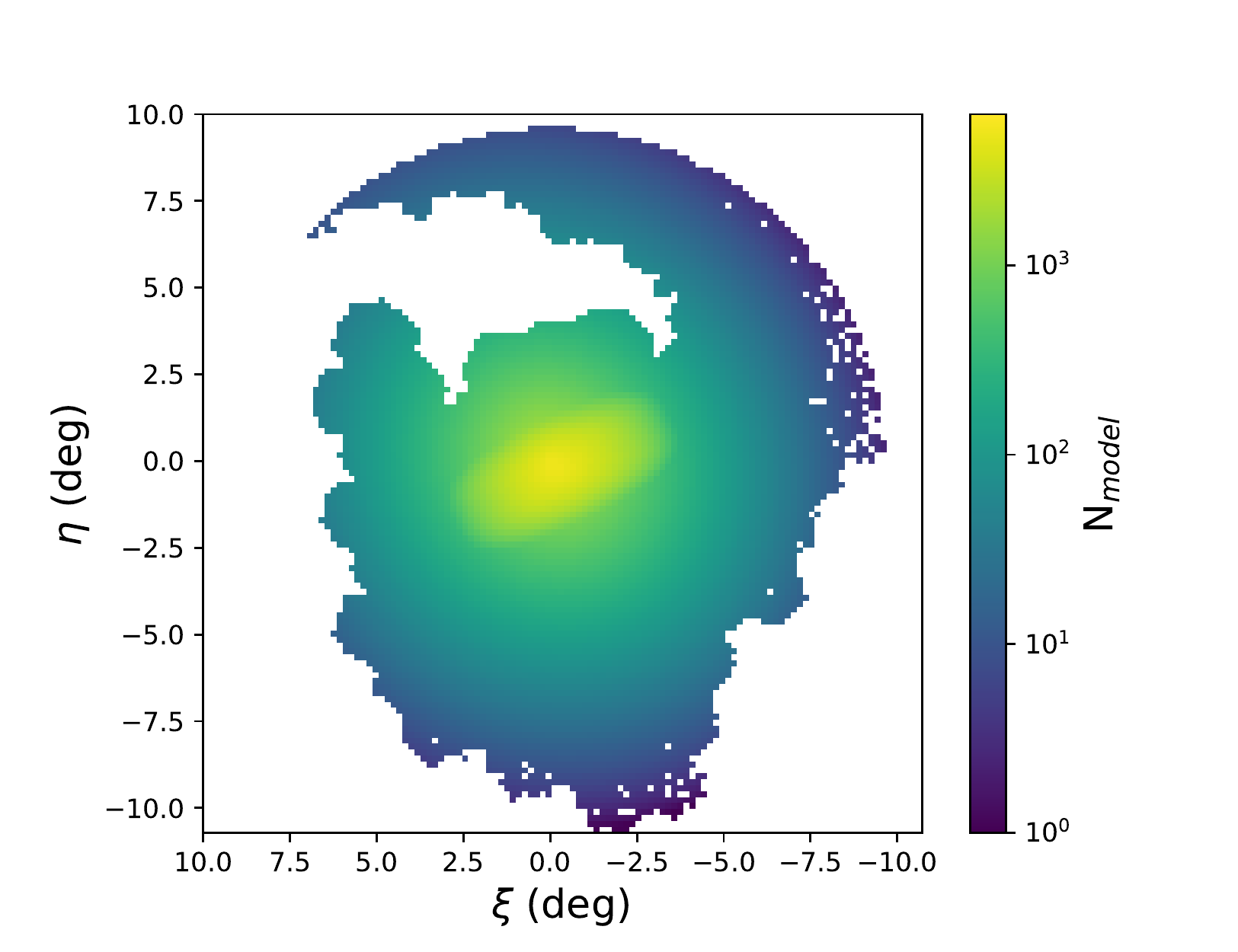}}}
\caption{1D and 2D Posterior probability distribution functions of the LMC disk and bar model fitting parameters. The dashed lines represent the 16th, 50th, and 84th percentiles of each 1D distribution. Contours enclose the top 68\%, 95\%, and 99.7\% of each 2D distribution. In the top right corner, we present the best-fit RC disk model. 
\label{diskmodel}}
\end{figure*}

\subsection{Modeling the Disk}\label{modeldisk}
We use only the RC stars, not the bright MS stars, as tracers of the LMC exponential disk and bar, as they sample a population that should be dynamically well mixed. We decompose the LMC into a disk and bar using a Bayesian technique with the Markov chain Monte Carlo (MCMC) sampler {\tt emcee}\footnote{\url{http://dan.iel.fm/emcee}} \citep{foremanmackey13}. With the Bayesian approach, we can extract the posterior probability distribution functions (pPDFs) of each of the model parameters, as well as the covariance between parameters. We can also incorporate prior knowledge about the model parameters. We use flat priors within physically meaningful ranges and zero outside those ranges (see Table~\ref{param_2comp}). 

Specifically, we impose a conservative prior on bar size, motivated by its apparent morphology and the model prediction \citep[$<$ 3.5\degr;][]{vandermarel01b,subramaniam04,zaritsky04b,besla12}. The bar size can be inferred from the disk geometry as well. In the inner disk where the bar dominates the disk morphology, the warped bar induces a larger inclination, and twists the position angle of the line of nodes with respect to the rest of the LMC disk. In particular, the position angle changes rapidly at 2\degr $< \rho <$ 3.5\degr\,and stays flat beyond that \citep[see Figure 13 in][]{choi18}, indicating the maximum length of the bar in the RC star count map is unlikely to be larger than 3.5\degr. \citet{vandermarel01b} reported a very similar behavior in their RGB star count map. Thus, we set the maximum bar size to be 3.5\degr. Furthermore, without setting this limit, the bar model always tries to fit substructures in the inner disk (e.g., sub-dominant spiral arm and even a part of the ring-like overdensity) as much as possible, which is not physically meaningful. 

We also impose a uniform prior on the boxiness (1.5 $< C_{0} <$ 3) to consider the range that roughly describes the stellar density contours of the LMC bar region (see the black contours in the left panel of Fig~\ref{rcmap}). The density contours of the bar region do not suggest significant deviation from a pure ellipse (i.e., $C_{0}$ = 2); the shape of the bar is neither disky nor very boxy. \citet{gadotti11} analyzed $\sim$300 barred galaxies in the local universe and found $C_{0}$ ranges between 1.5 and 3.5 with a peak at 3. \citet{kim15} characterized the shape of bars in 144 face-on barred galaxies, and found the maximum $C_{0}$ value of $\sim$3.5 in their sample. Our adopted prior range for $C_{0}$ is well within the range of values found for external barred galaxies.

We conduct a rigorous test on the effect of our choice of the prior ranges by modeling the disk with many different combinations of prior ranges on the bar size (3\degr $< r_{out,max} <$ 8\degr) and its boxiness (3 $< C_{0,max} <$ 8). The existence of the ring-like overdensity, its overall shape, and the maximum amplitude turn out to be robust against the variations in the bar parameter priors. The position angles of both the exponential disk and bar, and the bar ellipticity, are also insensitive to the prior ranges. The disk axis ratio and the scale length do not vary significantly either. If we allow a more generous range for $C_{0}$, the resulting bar always becomes very boxy ($C_{0} \gtrsim$ 3.5) to make up for non-bar structures, such as a sub-dominant arm emanating from the east end of the bar and a part of the ring-like overdensity regardless of the allowed range of the bar size. When keeping the same range for the bar size, the boxiness is the only model parameter to be significantly changed, while a more generous range for the bar size for a fixed prior range on $C_{0}$ returns an exponential disk with a larger scale length and a larger ellipticity that overpredicts the number of stars particularly in the outer disk. These larger disks with a higher ellipticity, however, are inconsistent with previous studies \citep[e.g.,][]{vandermarel01b,Saha2010,balbinot15,mackey16}.

To achieve the computational efficiency and accuracy, we compute the log probability:
\begin{equation} \label{eq:logbayes}
  lnP(\mu_{m} \midline \mu_{obs}) \propto lnP(\mu_{obs} \midline \mu_{m})\,+ lnP(\mu_{m}),
\end{equation}
where $lnP(\mu_{obs} \midline \mu_{m})$ is a log-likelihood and $lnP(\mu_{m})$ is the logarithm of the prior probability. For the stellar number count modeling, we use the Poisson likelihood: 
\begin{equation}
\begin{aligned}
lnP(\mu_{obs} \midline \mu_{m}) = \sum_i^{nx} \sum_j^{ny} &\big[\mu_{obs,ij}\,ln(\mu_{m,ij})\big. \\
&\big. - \mu_{m,ij} - ln(\mu_{obs,ij}!)\big].
\end{aligned}
\end{equation}

Instead of starting the walkers randomly sampled from flat priors, we run MCMC for 100--1000 steps several times to find a good initial guess for parameters. Restarting the walkers around the maximum log-probability position of the last step in the previous run is an efficient way to find reasonable initial guesses of the parameters. This technique also helps prevent the walkers from getting stuck at local likelihood maxima. 

Once we set a reasonable initial point in the parameter space, we start 100 independent walkers in a small $11$-dimensional sphere around the initial guess. At earlier steps in the chain, the walkers do not represent the true pPDFs. Once we run MCMC long enough \citep[$\sim$10 autocorrelation times;][]{foremanmackey13}, the ensemble of the walkers approximates the true pPDFs and thus we can estimate the parameters after discarding the burn-in steps in the chain. We determine conservatively the length of the burn-in phase based on the log probability over steps in the chain to assure the minimal effect of initial values.

\begin{figure}
\includegraphics[width=10cm,trim=1.5cm 1.5cm 0cm 2.5cm, clip=True]{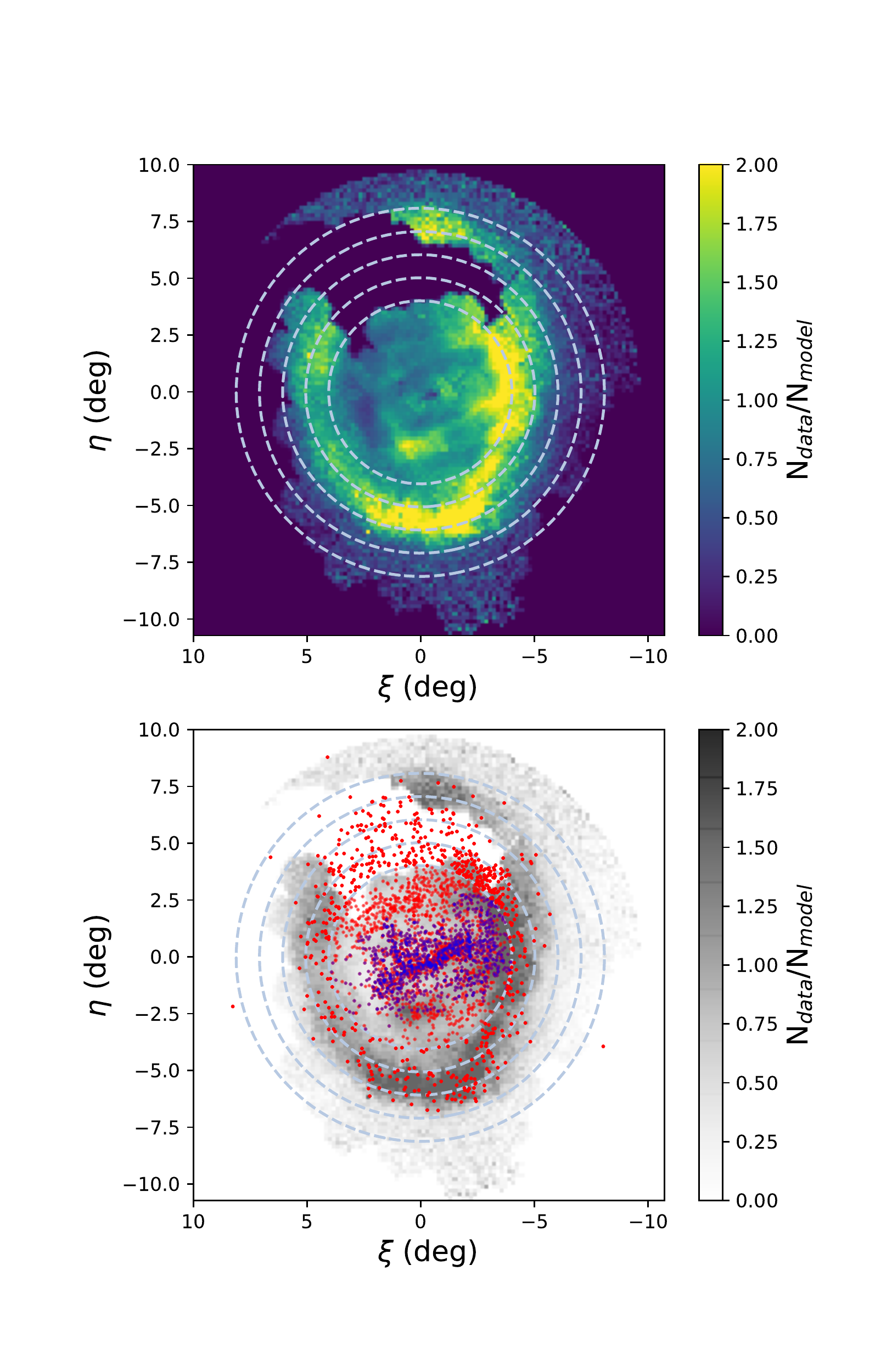}
\caption{Top: Residual density map (data/model) of the LMC disk.  The elliptical ``ring'' structure is clearly visible, circumscribing the LMC with a position angle coverage of $\sim$270\degr. Bottom: Intermediate-age star clusters from \citet{bica08} on top of the residual map (gray for clarity). The red dots are all star clusters from the \citet{bica08} catalog, ranging in age between a few Myr to 4~Gyr without specific ages for individual clusters. The blue dots are the clusters with ages available from \citet{nayak16}. Most of them are young ($<$ 250~Myr) star clusters overlapping with the star-forming bar. Outside the bar, there are a number of clusters that align with the ring feature. 
\label{scdist}}
\end{figure}

\section{Stellar Overdensity in the LMC Disk}
Figure 3 shows the one-dimensional (1D) marginalized posterior distributions and two-dimensional (2D) joint posterior contours for the four exponential disk parameters and seven bar parameters. The 2D contour plots show covariances between all pairs of parameters. Specifically, there are strong correlations between the central stellar number density ($\mu_{0,d}$) and disk scale length ($r_d$), between $r_d$ and disk axis ratio, between the bar central surface density $\mu_{0,b}$ and $\alpha$ and $\beta$, and between $\alpha$ and $\beta$. We list the resulting estimates for each parameter distribution in Table~\ref{param_2comp}.

With the exception of the two bar parameters ($r_{out}$ and $C_{0}$), all other parameters are well constrained. When the $r_{out}$ reaches to its maximum allowed value, the bar model tries to fit the surrounding stellar number density by becoming boxier. The main purpose of modeling the disk and bar simultaneously is to help characterize the amplitude of the ring-like overdensity better, not to model the bar perfectly. Furthermore, non-inclusion of a bar component in our disk fitting (i.e., exponential disk only modeling) does not change the existence, overall ring-like shape, and the maximum amplitude of the overdensity. However, the variation of the amplitude as a function of position angle depends on the inclusion of a bar component and the maximum allowed $r_{out}$, especially on the west and east sides where both the eastern and western ends of the bar cross the ring-like overdensity. The amplitude in the north and south is reasonably robust against the inclusion of a bar component.  

Our best-fit exponential disk parameters are consistent with previous studies. Specifically, the disk scale length of $\sim$1.67~kpc, the ellipticity of $\sim$0.2, and the disk position angle of $\sim$227\degr\,are consistent with the previous investigations of the LMC disk \citep[e.g.,][]{vandermarel01b}. This indicates that the major axis of the LMC disk elongates in the direction that is almost perpendicular to the line of nodes \citep[$\sim$150\degr;][]{choi18}. Although we limit the maximum bar size for physical reasons, we would not argue that we constrain the bar well. Broadly speaking, the bar is boxy and its major axis is almost parallel to the line of nodes, and its minor axis is slightly shorter than the half-length of its major axis. 

In the top panel of Figure~\ref{scdist}, we present the ratio map of the RC star count map over the disk model fit. It clearly reveals a ring-like stellar overdensity spanning all azimuthal angles at radii between 4\degr\,and 7\degr, depending on position angle, with a mean radius of $\sim$6\degr. The west and east sides are closer (4--5\degr) to the center, while the north and south sides are farther away (5--7\degr). We emphasize that there is no spatial overlap between the MS maps of stars younger (brighter) than 1~Gyr (19~mag) and the overdensity, indicating it is mostly composed of stars older than at least 1~Gyr. The amplitude of the overdensity also varies with position angle, and reaches up to 2.5$\times$ as large as the smooth underlying disk. The width of the overdensity is $\sim$1\degr\,and varies little with position angle. An excess of RGB and AGB stars in the radial stellar number density profile was also detected at $\sim$6\degr\,by \citet{vandermarel01b}. In addition, \citet{belokurov17} described the radial density profile of LMC RR Lyrae, which stems from older populations than the RC, as a broken power law with the break at $\sim$7\degr. Although the older RR Lyrae population might have a different density profile than the RC, we suggest that it is possible that the RR Lyrae profile could also be one single exponential with an overdensity at $\sim$6--7\degr\,making it appear as a broken power law (see the right panel of Figure~\ref{rcmap}). Indeed, a shallower radial profile was found in the LMC outer disk and periphery, which is a low-density envelope of stellar component extending to $\sim$22\degr, with a break at much larger radii of $\sim$13--15\degr\,\citep{nidever18}. \citet{majewski09} also suggested the break between the inner and outer density profiles is at $\sim$9\degr. Furthermore, \citet{Saha2010} probed to the north of the LMC out to $\sim$16\degr, and even found that their density profile is well described by a single exponential over the radial range explored. 

While the eastern side of the overdensity appears round, the western side is somewhat linear, with a kink at ($\xi$,$\eta$) = ($-$4\degr,$-$1\degr). This linearity is also seen in the star count map by \citet{devaucouleurs55}, \citet{devaucouleurs72}, \citet{helmi18}, and numerical models of the LMC \citep{besla16}. Asymmetry induced by repeated interactions with the SMC can naturally explain this linearity and kink in the overdensity \citep[e.g.,][]{yozin14,besla16}. 

An overdense structure seen at $\rho\sim$ 7.5\degr\,just north of the LMC center might not be a continuation of the ring-like overdensity in the southern disk. In fact, this structure coincides with the arc feature found by \citet{devaucouleurs72,besla16}, suggesting that this is a distinct substructure from our ring-like overdensity at 6\degr. Because of the gap in the northern disk between the SMASH and DES data, it is hard to tell whether the ring-like overdensity at 6\degr\,is a complete ring or a pseudo-ring from this data set. Regardless of its full shape, an almost ring-like spiral arm is a common feature found in highly perturbed spiral galaxies, such as is likely the case for the LMC. 

Another interesting feature, but much weaker than the ring overdensity, is seen in the ratio map at smaller radii. This structure emerges from the bottom of the eastern side of the bar, stretches toward the southwest, and finally merges into the ring-like overdensity at 6\degr. This feature is evident in the deep optical imaging presented in \citet{besla16} and denoted as a sub-dominant arm. 

\citet{mackey16} found a portion of a ring-like structure in the northern disk at radii between 9\degr\,and 13\degr, but their data were not sufficient to further investigate that structure at smaller radii. Our ring-like overdensity at 6\degr\,does not seem to be the continuation of the feature they detected at this large radius or their stream-like feature at even larger radius. However, both features could potentially have been created by the same mechanism as our overdensity, i.e., repeated tidal interactions with the SMC.

In the bottom panel of Figure~\ref{scdist}, we present the spatial distribution of star clusters on top of the ratio map. The ring-like overdensity at 6\degr\,seems to be consistent with the spatial distribution of intermediate-age star clusters \citep{bica08}. Unfortunately, specific ages of individual clusters are not provided by \citet{bica08}. \citet{nayak16} measured the ages of \citet{bica08} clusters around the bar and showed that most of them are younger than $<$ 250~Myr. If we exclude these younger clusters, the spatial correlation between the overdensity and the rest of \citet{bica08} clusters becomes more clear. A star cluster catalog with masses and ages compiled from the SMASH data will be available soon (L. C. Johnson et al., in preparation).

\section{Discussion}
There are two key properties of the ring-like overdensity at 6\degr\: (1) it mostly consists of stars older than $\sim$1~Gyr; and (2) the amplitude of the overdensity is up to 2.5$\times$ higher than the smooth underlying disk. Although our findings alone are not sufficient to distinguish between the two candidate scenarios discussed below, the detailed present-day morphology of the LMC can provide useful constraints on the recent dynamical evolution of the interacting LMC--SMC pair. This is because any characteristic lopsidedness seen the LMC is highly likely originated from repeated tidal interactions with the SMC \citep[e.g.,][]{athanassoula96,besla12,yozin14}. Here, we explore the two most possible scenarios that might explain the origin of this structure. 

\subsection{Completely Wrapped around Spiral Arm?} 
In the first scenario, the ring-like overdensity formed from the evolution of a one-armed spiral. A one-armed spiral, triggered by repeated tidal encounters with the SMC, can wind up around the LMC main body over time \citep[$\sim$1--2 Gyr, e.g.,][]{gomez13,besla16}. The amplitude of this structure may have become more pronounced after the SMC's recent collision \citep[$\sim$150 Myr ago,][]{zivick18}. This picture is consistent with simulations of repeated interactions between the MCs \citep{besla12}. Theoretical simulations of this general scenario have illustrated that a tidally induced ring-like spiral overdensity can be 2--3$\times$ larger than the smooth disk \citep[e.g.,][]{purcell11,gomez13}, which agrees with the amplitude of the observed structure. 

In this scenario, the lack of young stellar populations can be naturally understood in that the disturbances induced by the SMC at earlier times primarily affect the outer disk, which is dominated by older stars and likely has no sufficient dense gas to form stars. Even if star formation was accompanied with the formation of a spiral arm by earlier (more than 1~Gyr ago) encounters, the continuation of star formation to the present day might be difficult owing to the very low gas density in the outskirts of the LMC disk. The mean \hi\,column density decreases rapidly beyond 2.5~kpc and drops below $N$(\hi) = 10$^{20}$~cm$^{-2}$ beyond 5~kpc \citep{stevelysmith03}, indicating that the outer LMC disk is surrounded by very low density \hi\,gas. There are only six young stars identified in the outskirts of the LMC \citep{monibidin17}, clearly showing that stars barely form out of this low gas density. The absence of young MS stars also has been reported in other galaxies' outer disks where the gas density is lower than 10$^{20}$~cm$^{-2}$ ($=$ 1~\msun~$pc^{-2}$) \citep[e.g.,][]{grossi11,radburnsmith12,hunter13}. Furthermore, the star formation histories measured in the LMC disk exhibit outside-in quenching of star formation; star formation activity has propagated inward over the last $\sim$1~Gyr \citep{Meschin2014}. This suggests that the LMC gas disk has significantly shrunk in the past gigayear. Secular gas depletion causes contraction of the star-forming disk over time in most dwarf galaxies \citep[e.g.,][]{stinson09}, and thus can be expected in the LMC as well. 

However, external processes such as ram-pressure-stripping and tidal interaction with the SMC might be also important gas-stripping mechanisms for the LMC outskirts. The LMC gas disk has experienced ram-pressure-stripping since the MCs first entered into the virial radius of the MW \citep[$\sim$1~Gyr ago;][]{besla12}. The sharp falloff of the gas density beyond 4~kpc in the northeast leading edge, which is the moving direction of the LMC, with an extended old stellar disk well beyond $\sim$5~kpc is a clear evidence of ram-pressure-stripping \citep{salem15}. Repeated LMC--SMC interactions may strip gas from the LMC as well as the SMC \citep[e.g.,][]{nidever08, richter13, pardy18}. Thus, it is likely that the combination of all these processes significantly decreased the gas density in the outskirts over the last 1--2~Gyr, and the resulting low gas density prevented density waves (if the ones that are associated with the overdensity exist) from triggering shock-induced star formation in the overdensity during the past 1~Gyr. Or the strength of a shock could be too weak to form molecular clouds by compressing diffuse gas entering into a spiral arm.

\subsection{Collisionally Induced Ring?}
In the second scenario, the ring-like overdensity is an immediate tidal response to the recent collision with the SMC \citep[e.g., See Figure 3 in][]{athanassoula96}. In this scenario, the structure can be interpreted as `ringing' \citep[e.g.,][]{gomez12b,gomez12a}. \citet{gomez13} showed that radial density waves can induce a significant overdensity when a satellite galaxy plunges through the disk of its host galaxy. Although their simulation was optimized for the interaction between the MW and the Sagittarius dwarf spheroidal galaxy (with roughly a 1:10 mass ratio), the same logic can be applied to the LMC and SMC interaction, which have a similar mass ratio. Indeed, pronounced disturbances are expected in the LMC disk after the recent SMC collision \citep[e.g.,][]{besla12, pardy16}. In this scenario, recent star formation may be expected in the ring-like overdensity. However, there is no evidence of either ongoing star formation traced by H$\alpha$ emission \citep[SHASSA;][]{gaustad01} or recent star formation traced by UV \citep[GALEX;][]{martin05} and far-IR \citep[AKARI;][]{murakami07} in the ring-like structure. If the overdensity is a simple stellar response to the density waves, the age of the composing stars is not necessarily connected to the age of the structure. The fact that the overdensity only shows up in intermediate-age or older stars may simply indicate that there were no dense gas to form new stars or no pre-existing young stars at those radii at the time of the recent collision. Due to the gas removal processes discussed in the previous section, it is likely that there was actually no sufficient dense gas left to form stars in the outer disk by the time of the recent collision ($\sim$150~Myr ago).

\subsection{Constraints on the LMC--SMC Evolution}
In both scenarios, the resulting structures are kinematically induced and thus expected to be short-lived \citep[$\sim$1--2 Gyr;][]{berentzen03,yozin14,pardy16}, although repetitive encounters can reform the structure. Furthermore, in both scenarios, the origin of the ring-like overdensity at 6\degr\ is a product of tidal interactions with the SMC, rather than the MW. To test the origin of this structure, additional observational and numerical studies should be conducted. Observationally, we can look for: (1) radial propagation either in the stellar motions themselves or older star formation across the ring overdensity as in \citet{choi15}; and (2) radial variations of the mean vertical stellar velocity, which is expected in vertical density waves \citep{gomez16,gomez17}. If one detects radial propagation in the stellar motions or radial variations in the vertical stellar velocity, this would indicate that the overdensity is likely a product of the recent collision. Numerically, the existence of the ring-like overdensity and its properties (position, amplitude, and shape) can be tested in high-resolution simulations by investigating different values of the LMC--SMC mass ratio, impact parameter, and timing of the collision. Because the position and amplitude of the ring-like overdensity strongly depend on these parameters, and in particular its impact parameter with the LMC, our findings will strongly constrain the recent ($<$ 500~Myr) dynamical evolution of the MCs. Given the sensitivity of the interaction history of the MCs to the tidal field of the MW, by including their kinematic information \citep[e.g.,][]{helmi18,niederhofer18,zivick18} we can also constrain the total dark matter halo mass of the MW -- a low-impact-parameter collision between the MCs is less likely as the mass of the MW increases \citep{zivick18}. Thus, searching through the parameter space to simultaneously reproduce the observed MCs' morphology, kinematics, and its large-scale gas structure will constrain both dynamical models for the MCs and the halo mass of the MW. 

Finally, these advances will help improve our general understanding of galaxy dynamics and the morphological evolution of interacting dwarf galaxies, using the MCs as a prototype for an interacting/colliding pair of galaxies. Dwarf--dwarf interactions are ubiquitous, but the morphological consequence of mergers between two small galaxies has not been fully investigated. Because the MCs' dynamics are complicated by their proximity to the MW, comparison with dwarf pairs in the field will enable us to better disentangle the environmental influence of a massive host vs. dwarf--dwarf tidal effects \citep{stierwalt15,pearson16}. LMC--SMC analogs are rare in the field \citep[$<$ 1\%;][]{besla18}, but nearby examples exist. Magellanic Irregulars, NGC 4027 and NGC 3664, each have an SMC-mass companion. Both NGC 4027 \citep{phookun92} and NGC 3664 \citep{wilcots04} have one prominent spiral, an off-centered bar, and a smaller companion at $\sim$25-30~kpc from them. Even a sub-dominant spiral feature was detected in NGC 4027 \citep{phookun92}. Despite all these similarities to the LMC, there are two main differences: their completely wrapped ring contains both young and old stars; and their gas disks extend well beyond their stellar disks. This might indicate that the ring-like structure can be an evolutionary feature in interacting Magellanic-type galaxies.

\acknowledgements
The authors thank Facundo G\'omez for useful discussions on the origin of the ring-like overdensity feature. We are also grateful to the referee for providing helpful comments to improve the paper. Y.C. and E.F.B. acknowledge support from NSF grant AST 1655677. M.-R.L.C. acknowledges support from the European Research Council (ERC) under the European Union's Horizon 2020 research and innovation program (grant agreement No. 682115). T.D.B. acknowledges support from the European Research Council (ERC StG-335936). A.M. acknowledges partial support from CONICYT FONDECYT regular 1181797. D.M.D. acknowledges support by Sonderforschungsbereich (SFB) 881 ``The Milky Way System'' of the German Research Foundation (DFG), particularly through subprojects A2. This research uses services or data provided by the NOAO Data Lab. NOAO is operated by the Association of Universities for Research in Astronomy (AURA), Inc. under a cooperative agreement with the National Science Foundation.

This work is based on observations at Cerro Tololo Inter-American Observatory, National Optical Astronomy Observatory (NOAO Prop. IDs: 2013A-0411 and 2013B-0440; PI: Nidever), which is operated by the Association of Universities for Research in Astronomy (AURA) under a cooperative agreement with the National Science Foundation. This project used data obtained with the Dark Energy Camera (DECam), which was constructed by the Dark Energy Survey (DES) collaboration. Funding for the DES Projects has been provided by the U.S. Department of Energy, the U.S. National Science Foundation, the Ministry of Science and Education of Spain, the Science and Technology Facilities Council of the United Kingdom, the Higher Education Funding Council for England, the National Center for Supercomputing Applications at the University of Illinois at Urbana-Champaign, the Kavli Institute of Cosmological Physics at the University of Chicago, Center for Cosmology and Astro-Particle Physics at the Ohio State University, the Mitchell Institute for Fundamental Physics and Astronomy at Texas A\&M University, Financiadora de Estudos e Projetos, Funda\c{c}\~ao Carlos Chagas Filho de Amparo, Financiadora de Estudos e Projetos, Funda\c{c}\~ao Carlos Chagas Filho de Amparo \`a Pesquisa do Estado do Rio de Janeiro, Conselho Nacional de Desenvolvimento Cientifico e Tecnol\'ogico and the Minist\'erio da Ci\^encia, Tecnologia e Inova\c{c}\~ao, the Deutsche Forschungsgemeinschaft, and the Collaborating Institutions in the Dark Energy Survey. The Collaborating Institutions are Argonne National Laboratory, the University of California at Santa Cruz, the University of Cambridge, Centro de Investigaciones En\'ergeticas, Medioambientales y Tecnol\'ogicas-Madrid, the University of Chicago, University College London, the DES-Brazil Consortium, the University of Edinburgh, the Eidgen\"ossische Technische Hochschule (ETH) Z\"urich, Fermi National Accelerator Laboratory, the University of Illinois at Urbana-Champaign, the Institut de Ci\`encies de l'Espai (IEEC/CSIC), the Institut de F\'isica d'Altes Energies, Lawrence Berkeley National Laboratory, the Ludwig-Maximilians Universit\"at M\"unchen and the associated Excellence Cluster Universe, the University of Michigan, the National Optical Astronomy Observatory, the University of Nottingham, the Ohio State University, the University of Pennsylvania, the University of Portsmouth, SLAC National Accelerator Laboratory, Stanford University, the University of Sussex, and Texas A\&M University.

\facility{Blanco (DECam).}

\software{scipy \citep{jones01}, numpy \citep{vanderwalt11}, matplotlib \citep{hunter07}, ipython \citep{PER-GRA:2007}, astropy \citep{astropy18}, emcee \citep{foremanmackey13}, and corner \citep{corner}.}

\bibliographystyle{aasjournal}
\bibliography{LMCdisk_refs}
\clearpage

\end{document}